\documentclass[prl,amssymb,amsmath,amsfonts,floatfix,twocolumn]{revtex4}

\begin{document}

\title{Scalars, Vectors and Tensors from Metric-Affine Gravity}

\author{Canan N. Karahan, Asl{\i} Alta{\c s}, Durmu{\c s} A. Demir}

\affiliation{Department of Physics, Izmir Institute of
Technology, Izmir, TR35430, Turkey}

\begin{abstract}

The metric-affine gravity provides a useful framework for
analyzing gravitational dynamics since it treats metric
tensor and affine connection as fundamentally independent
variables. In this work, we show that, a metric-affine gravity
theory composed of the invariants formed from non-metricity, torsion and
curvature tensors can be decomposed into a theory of scalar, vector and tensor
fields. These fields are natural candidates for the ones needed by various
cosmological and other phenomena. Indeed, we show that the model accommodates TeVeS
gravity (relativistic modified gravity theory), vector inflation, and
aether-like models. Detailed analyses of these and other phenomena can lead
to a standard metric-affine gravity model encoding scalars, vectors and
tensors.
\end{abstract}
\maketitle

\section{Introduction}

Spacetime is a smooth manifold $M(g;\between)$ endowed with a metric $g$ and connection $\between$. Metric is responsible
for measuring the distances while affine connection governs the straightness of curves and twirling of the manifold. These two geometrical structures, the metric and connection, are fundamentally independent geometrical variables, and
they play completely different roles in spacetime dynamics.
If they are to exhibit any relationship it derives from dynamical equations a
\emph{posteriori}. This fact gives rise to an
alternative approach to Einstein's standard theory of general relativity: Metric-Affine Gravity.

The standard theory of general relativity is a purely metric theory of gravity since connection is completely
determined by the metric and its partial derivatives,  \emph{a priori}. This determination is encoded in the Levi-Civita
connection,
\begin{eqnarray}
\label{levi-civita}
    \Gamma^{\lambda}_{\alpha\beta} =\frac{1}{2} g^{\lambda\rho} \left(\partial_{\alpha} g_{\beta \rho} +
    \partial_{\beta} g_{\rho\alpha} - \partial_{\rho} g_{\alpha\beta}\right)
\end{eqnarray}
which defines a metric-compatible covariant derivative \cite{SC}.

The metric-affine theory of gravity (similar to the Palatini formalism \cite{palatini} in philosophy),
which treats an metric tensor and connection as independent variables \cite{SC,PP},
encodes a more general approach to gravitation by breaking up the a priori
relation (\ref{levi-civita}). This breaking inherently reveals the new dynamic structures
torsion, nonmetricity in addition to curvature.

In this work we shall study metric-affine gravity in regard to decomposing the affine connection
into independent vectors, tensors and scalars. We shall, in particular, be able to derive certain
interactions using solely the {\it geometrical} sector with no reference to the matter sector that
contains the known forces and species. Our starting point will be the fundamental independence of
connection and metric, and the field content of the connection in the most general case.

The outline of the paper is as follows. In Sec. II below we first construct the most general `connection'
involving physically `distinct and independent' structures, and then form a general action containing
vector and tensor fields. In Sec. III we give specific applications of the derived action to vector inflation and
TeVeS theory. Here we also discuss the relation of the model to the ones in the literature. In Sec. IV we conclude.

\section{Tensor-Vector Theories from Non-Riemannian Geometry}

An affine connection, whose components to be symbolized by ${\between}^{\lambda}_{\alpha
\beta}$, governs parallel transport of tensor fields along a given
curve in spacetime, and parallel transport around a closed curve,
after one complete cycle, results in a finite mismatch if the spacetime is curved.
Curving is uniquely determined by the Riemann curvature tensor
\begin{eqnarray}
\label{riemann}
\mathbb{R}^{\mu}_{\alpha \nu \beta}\left(\between\right) = \partial_{\nu} \between^{\mu}_{\beta\alpha}
-\partial_{\beta}\between^{\mu}_{\nu \alpha} + \between^{\mu}_{\nu \lambda} \between^{\lambda}_{\beta \alpha} -
\between^{\mu}_{\beta \lambda} \between^{\lambda}_{\nu \alpha}
\end{eqnarray}
which is a tensor field made up solely of the non-tensorial
objects $\between^{\lambda}_{\alpha \beta}$ and their partial
derivatives. Notably, higher rank tensors involving $(n+1)$ partial
derivatives of $\between^{\lambda}_{\alpha \beta}$ are given by $n$-th covariant
derivatives of $\mathbb{R}^{\mu}_{\alpha \nu \beta}$, and hence,
$\mathbb{R}^{\mu}_{\alpha \nu \beta}$ acts as the seed tensor
field for a complete determination of the spacetime curvature.

Affine connection determines not only the curving but also the
twirling of the spacetime. This effect is encoded in the torsion tensor
\begin{eqnarray}
\label{torsion}
\mathbb{S}^{\lambda}_{\alpha \beta}\left(\between\right) =   \between^{\lambda}_{\alpha\beta} -
\between^{\lambda}_{\beta\alpha}
\end{eqnarray}
which participates in structuring of the spacetime together
with curvature tensor. Torsion vanishes in geometries with
symmetric connection coefficients, $\between^{\lambda}_{\alpha\beta} = \between^{\lambda}_{\beta\alpha}$.

The spacetime gets further structured by the notions of distance and
angle if it is endowed with a metric tensor $g_{\alpha \beta}$ comprising
clocks and rulers needed to make measurements. The connection coefficients and
metric tensor are fundamentally independent quantities. They exhibit no {\it a
priori} known relationship, and if they are to have any it must
derive from some additional constraints. This property is best expressed by
the non-metricity tensor
\begin{eqnarray}
\label{nonmetricity}
\mathbb{Q}^{\alpha\beta}_{\lambda}\left(g,\between\right) = \nabla^{\between}_{\lambda} g^{\alpha\beta}
\end{eqnarray}
which is non-vanishing for a general connection $\between^{\lambda}_{\alpha\beta}$. This
rank (2,1) tensor would identically vanish if the connection were compatible with the
metric. Indeed, in GR, for instance, the constraint to relate $\between^{\lambda}_{\alpha\beta}$
to $g_{\alpha\beta}$ is realized by imposing $\between^{\lambda}_{\alpha\beta}=\Gamma^{\lambda}_{\alpha\beta}$
from the scratch, where $\Gamma^{\lambda}_{\alpha\beta}$ is the Levi-Civita connection (\ref{levi-civita})
which respect to which metric stays covariantly constant, $\nabla_{\lambda}^{\Gamma} g_{\alpha\beta} = 0$, and
hence, non-metricity vanishes identically. Furthermore, for this particular connection,
the torsion also vanishes identically since $\Gamma^{\lambda}_{\alpha\beta}= \Gamma^{\lambda}_{\beta\alpha}$
by definition.

The curving and twirling of the spacetime are governed by the connection $\between^{\lambda}_{\alpha\beta}$. The
metric
tensor has nothing to do with them, and the Riemann curvature tensor (\ref{riemann}) contracts, with
no involvement of the metric tensor,  in three different ways to generate the associated
Ricci tensors of $\between^{\lambda}_{\alpha\beta}$:
\begin{itemize}
\item $\mathbb{R}_{\alpha\beta}\left(\between\right) \equiv
    \mathbb{R}^{\mu}_{\alpha\mu\beta}\left(\between\right)$\,,
\item $\widehat{\mathbb{R}}_{\alpha\beta}\left(\between\right) \equiv
    \mathbb{R}^{\mu}_{\alpha\beta\mu}\left(\between\right) = - \mathbb{R}_{\alpha\beta}\left(\between\right)$\,,
\item $\overline{\mathbb{R}}_{\alpha\beta}\left(\between\right) \equiv
    \mathbb{R}^{\mu}_{\mu\alpha\beta}\left(\between\right) = \partial_{\alpha} \between^{\mu}_{\beta \mu} -
    \partial_{\beta} \between^{\mu}_{\alpha \mu}$\,.
\end{itemize}
The reason for having more than one Ricci tensor is that the Riemann tensor (\ref{riemann}) possesses only a single
symmetry $ \mathbb{R}^{\mu}_{\alpha \nu \beta}\left(\between\right) = -\mathbb{R}^{\mu}_{\alpha \beta
\nu}\left(\between\right)$. It is this symmetry property that gives
the relation $\widehat{\mathbb{R}}_{\alpha\beta}\left(\between\right) = -
\mathbb{R}_{\alpha\beta}\left(\between\right)$ between the first two Ricci tensors above. The third Ricci tensor
$\overline{\mathbb{R}}_{\alpha\beta}\left(\between\right)$ does not exist in the General Relativity (GR) since symmetries of the Riemann
tensor, $\mathbb{R}_{\mu\alpha\nu\beta}\left(\Gamma\right) \equiv g_{\mu \mu^{\prime}}
\mathbb{R}^{\mu^{\prime}}_{\alpha\nu\beta}\left(\Gamma\right) = - \mathbb{R}_{\mu\alpha\beta\nu}\left(\Gamma\right) = -
\mathbb{R}_{\alpha\mu\nu\beta}\left(\Gamma\right) = \mathbb{R}_{\nu \beta \mu\alpha}\left(\Gamma\right)$, admits
only one single independent Ricci tensor, the $\mathbb{R}_{\alpha\beta}\left(\Gamma\right)$ defined above.

Unlike the Riemann and Ricci tensors, the curvature scalar is obtained only by contraction with the inverse metric.
Therefore, one finds the curvature scalar
\begin{eqnarray}
\mathbb{R}\left(g,\between\right) \equiv g^{\alpha\beta} \mathbb{R}_{\alpha\beta}\left(\between\right)  = -
g^{\alpha\beta} \widehat{\mathbb{R}}_{\alpha\beta}\left(\between\right) \equiv -
\widehat{\mathbb{R}}\left(g,\between\right)
\end{eqnarray}
from the first two Ricci tensors listed above. Likewise, the third Ricci tensor contracts to
\begin{eqnarray}
\overline{\mathbb{R}}\left(g,\between\right) = g^{\alpha \beta}
\overline{\mathbb{R}}_{\alpha\beta}\left(\between\right)  = 0
\end{eqnarray}
as dictated by the anti-symmetric nature of $\overline{\mathbb{R}}_{\alpha\beta}\left(\between\right)$. As a result,
the theory possesses two distinct Ricci tensors but a single Ricci scalar.

The action density describing matter and gravity is formed by invariants generated by the tensor fields above plus the
matter Lagrangian. A partial list includes
\begin{eqnarray}
\label{list}
&&\mathbb{R},\; \mathbb{S}\bullet\mathbb{S},\; \mathbb{Q}\bullet\mathbb{Q},\;
\mathbb{Q}\bullet\mathbb{S},\;\nonumber\\
&& \mathbb{R}^2,\; \mathbb{R} \bullet \mathbb{R},\; \overline{\mathbb{R}}\bullet \overline{\mathbb{R}},\; \nonumber\\
&& \mathbb{R}\bullet\mathbb{S}\bullet\mathbb{S},\; \mathbb{R}\bullet\mathbb{Q}\bullet\mathbb{Q},\;
\mathbb{R}\bullet\mathbb{Q}\bullet\mathbb{S},\;\nonumber\\
&& \overline{\mathbb{R}}\bullet\mathbb{S}\bullet\mathbb{S},\;
\overline{\mathbb{R}}\bullet\mathbb{Q}\bullet\mathbb{Q},\;
\overline{\mathbb{R}}\bullet\mathbb{Q}\bullet\mathbb{S},\;\nonumber\\
&&\mathbb{S}\bullet\mathbb{S}\bullet\mathbb{S}\bullet\mathbb{S},\;
\mathbb{Q}\bullet\mathbb{Q}\bullet\mathbb{Q}\bullet\mathbb{Q},\;\nonumber\\ &&
\mathbb{S}\bullet\mathbb{Q}\bullet\mathbb{Q}\bullet\mathbb{Q},\;
\mathbb{S}\bullet\mathbb{S}\bullet\mathbb{Q}\bullet\mathbb{Q},\; \nonumber\\
&&\mathbb{S}\bullet\mathbb{S}\bullet\mathbb{S}\bullet\mathbb{Q},\;
\mathcal{L}_{\texttt{matter}}\left(g,\between,\psi\right)
\end{eqnarray}
where ${\mathcal{L}_{m}\left(g,\between,\psi\right)}$ is the matter Lagrangian which explicitly involves the matter and
radiation fields $\psi$, the metric $g$ and the connection $\between$.
The first line of the list consists of mass dimension-2 invariants while the rest involve mass dimension-4 ones. Those
structures having mass dimension-5 or higher
are not shown. Also not shown are the invariants involving the covariant derivatives of the tensors.  The bullet (
$\bullet$ ) stands for contraction of the tensors in all possible ways by  using the metric tensor, in case needed.

The scalars in (\ref{list}), most of which do not exist at all in the GR, contain novel degrees of
freedom reflecting the non-Riemannian nature of the underlying geometry. These degrees of freedom
can be explicated via the decomposition of the connection
\begin{eqnarray}
\label{decomp}
\between^{\lambda}_{\alpha\beta} = \Gamma^{\lambda}_{\alpha\beta} + \Delta^{\lambda}_{\alpha\beta}
\end{eqnarray}
with respect to the Levi-Civita connection (\ref{levi-civita}), which is the most natural connection
one would consider in the presence of the metric tensor. In this decomposition, $\Delta^{\lambda}_{\alpha\beta}$,
being the difference between two connections, is a rank (1,2) tensor field, and it is the source of various
non-Riemannian invariants listed in (\ref{list}). To this end, in response to (\ref{decomp}),
the Ricci curvature tensor $\mathbb{R}_{\alpha\beta}\left(\between\right)$ splits as
\begin{eqnarray}
\label{decomp-R}
\mathbb{R}_{\alpha\beta}\left(\between\right) = R_{\alpha\beta}\left(\Gamma\right) +
\mathcal{R}_{\alpha\beta}\left(\Delta\right)
\end{eqnarray}
where $R_{\alpha\beta}\left(\Gamma\right)\equiv \mathbb{R}_{\alpha\beta}\left(\Gamma\right)$ is the Ricci curvature
tensor of the Levi-Civita connection, and
\begin{eqnarray}
\label{curve-extra}
\mathcal{R}_{\alpha\beta} = \nabla_{\mu} \Delta^{\mu}_{\beta\alpha} -\nabla_{\beta}\Delta^{\mu}_{\mu \alpha} +
\Delta^{\mu}_{\mu \nu} \Delta^{\nu}_{\beta \alpha} - \Delta^{\mu}_{\beta \nu} \Delta^{\nu}_{\mu \alpha}
\end{eqnarray}
where $\nabla_{\alpha}\equiv \nabla_{\alpha}^{\Gamma}$ is the covariant derivative of the Levi-Civita connection
$\Gamma^{\lambda}_{\alpha\beta}$.
This tensor is a rank (0,2) tensor field generated by the tensorial connection $\Delta^{\lambda}_{\alpha\beta}$ alone.
It is actually not a true curvature
tensor as it is generated by none of the covariant derivatives $\nabla^{\between}$ or $\nabla^{\Gamma}$. It is a
`quasi' curvature tensor.

In response to (\ref{decomp}), the purely non-Riemannian Ricci tensor
$\overline{\mathbb{R}}_{\alpha\beta}\left(\between\right)$ takes the form
\begin{eqnarray}
\label{Rnew-acik}
\overline{\mathbb{R}}_{\alpha\beta}\left(\between\right) &=& \partial_{\alpha} {\texttt{V}}_{\beta} - \partial_{\beta}
{\texttt{V}}_{\alpha}
= \nabla_{\alpha}^{\Gamma} {\texttt{V}}_{\beta} - \nabla_{\beta}^{\Gamma} {\texttt{V}}_{\alpha}
\end{eqnarray}
wherein the second equality, which ensures that $\mathbb{R}_{\alpha\beta}\left(\between\right)$
is a rank (0,2) anti-symmetric tensor field, follows from the symmetric nature of the
Levi-Civita connection, $\Gamma^{\lambda}_{\alpha\beta} = \Gamma^{\lambda}_{\beta\alpha}$. It is
obvious that $\overline{\mathbb{R}}_{\alpha\beta}\left(\between\right)$, in the form (\ref{Rnew-acik}),
is nothing but the field strength tensor
\begin{eqnarray}
\label{fst-V}
\overline{\mathbb{R}}_{\alpha\beta}\left(\between\right) \equiv \texttt{V}^{(-)}_{\alpha\beta}\equiv \partial_{\alpha}
{\texttt{V}}_{\beta} - \partial_{\beta} {\texttt{V}}_{\alpha}
\end{eqnarray}
of the Abelian vector
\begin{eqnarray}
\label{Vvec}
{\texttt{V}}_{\alpha} = \Delta^{\mu}_{\alpha \mu}
\end{eqnarray}
which is of purely geometrical origin. Consequently, purely non-Riemannian curvature tensor
$\overline{\mathbb{R}}_{\alpha\beta}\left(\between\right)$
plays a strikingly different role compared to $\mathbb{R}_{\alpha\beta}\left(\between\right)$ in that it directly
extracts a vector
field out of the underlying geometry.

As a result of (\ref{decomp}), the torsion and non-metricity tensors
\begin{eqnarray}
\label{torsion-acik}
\mathbb{S}^{\lambda}_{\alpha \beta}\left(\between\right) &=&   \Delta^{\lambda}_{\alpha\beta} -
\Delta^{\lambda}_{\beta\alpha}\\
\label{nonmetricity-acik}
\mathbb{Q}^{\alpha\beta}_{\lambda}\left(g,\between\right) &=&  \Delta^{\alpha}_{\lambda\mu} g^{\mu\beta} +
\Delta^{\beta}_{\lambda\mu} g^{\alpha\mu}
\end{eqnarray}
reduce to plain algebraic expressions in terms of $\Delta^{\lambda}_{\alpha\beta}$.

Having explicated the $\Delta^{\lambda}_{\alpha\beta}$ dependencies of the fundamental tensor
fields, it is time to ask what the tensorial connection actually is and what information about the
geometry can be extracted from it. In other words, $\Delta^{\lambda}_{\alpha\beta}$, which embodies
non-Riemannian ingredients of the underlying geometry, must be refined in order to extract the novel
geometrical degrees of freedom it contains. As the first option to think of, it is possible that there
exist a fundamental rank (1,2) tensor field $\delta^{\lambda}_{\alpha\beta}$, and the connection
$\Delta^{\lambda}_{\alpha\beta}$
equals just this fundamental tensor field. Though this is possible, at present there is no indication for
such higher spin fields, and thus, it is convenient to leave this possibility aside. The other option to
think of is that  $\Delta^{\lambda}_{\alpha\beta}$ could be made up of lower spin fields, {\it i. e.} vectors,
spinors and scalars. To this end, given its rank (1,2) nature, it is obvious that the tensorial connection
must be decomposable into vector fields, which might be fundamental fields or composites formed out of spinors
or scalars. In general, $\Delta^{\lambda}_{\alpha \beta}$ possesses 64 independent elements, and hence,
it should be fully parameterizable by 3 independent vector fields, whose nature will be further analyzed
in the sequel. One of the vectors is already defined by the contraction $\texttt{V}_{\alpha}$ in (\ref{Vvec}). The
other two
\begin{eqnarray}
\label{Uvec}{\texttt{U}}_{\alpha} &=& \Delta^{\mu}_{\mu \alpha}
\end{eqnarray}
and
\begin{eqnarray}
\label{Wvec}{\texttt{W}}^{\alpha} &=& g^{\mu \nu} \Delta^{\alpha}_{\mu \nu}
\end{eqnarray}
are conveniently defined through the remaining two distinct contractions of $\Delta^{\lambda}_{\alpha \beta}$. These
two vectors, unlike $\texttt{V}_{\alpha}$, do not possess an immediate kinetic term,
and if they are to have any, it must come from the invariants involving the gradients
of the fundamental tensors in (\ref{list}).

Given the metric tensor $g_{\alpha\beta}$, $\texttt{V}_{\alpha}$ in  (\ref{Vvec}), $\texttt{U}_{\alpha}$ in
(\ref{Uvec}), and
$\texttt{W}_{\alpha}$ in (\ref{Wvec}), the tensorial connection $\Delta^{\lambda}_{\alpha\beta}$
can be algebraically decomposed as
\begin{eqnarray}
\label{delta-decomp}
\Delta^{\lambda}_{\alpha\beta} &=& \delta^{\lambda}_{\alpha\beta} + a_v \texttt{V}^{\lambda} g_{\alpha\beta} + b_v
\texttt{V}_{\alpha} \delta^{\lambda}_{\beta} + c_v \delta^{\lambda}_{\alpha} \texttt{V}_{\beta} \nonumber\\
&+& a_u \texttt{U}^{\lambda} g_{\alpha\beta}  + b_u \texttt{U}_{\alpha} \delta^{\lambda}_{\beta} + c_u
\delta^{\lambda}_{\alpha} \texttt{U}_{\beta}\nonumber\\
&+& a_w \texttt{W}^{\lambda} g_{\alpha\beta}  + b_w \texttt{W}_{\alpha} \delta^{\lambda}_{\beta} + c_w
\delta^{\lambda}_{\alpha} \texttt{W}_{\beta}\nonumber\\
&+& \frac{1}{M^{2}} \sum \left( \nu_{x y} \texttt{V}^{\lambda} +  \upsilon_{x y} \texttt{U}^{\lambda} + \omega_{x y}
\texttt{W}^{\lambda}\right) \texttt{X}_{\alpha} \texttt{Y}_{\beta}
\end{eqnarray}
because of its higher spin assuming that a fundamental rank (1,2) tensor field $\delta^{\lambda}_{\alpha\beta}$ does
not exist at all. The sum in the last line runs over $\texttt{X}, \texttt{Y} = \texttt{V}, \texttt{U}, \texttt{W}$, and
$M$ is a mass scale expected to be around the fundamental scale of gravity, $M_{Pl}$. The decomposition necessarily
involves linear and trilinear combinations of the vectors. There cannot exist any other acceptable combinations of the
vectors. The expansion is unique in structure. However, one notices that all three defining relations (\ref{Vvec}),
(\ref{Uvec}), (\ref{Wvec}) are algebraic in nature, and thus, the dimensionless coefficients  $a$'s, $\dots$,
$\omega$'s cannot be prohibited to involve dressing factors of the form $\texttt{I}^{\delta}/M^{\delta}$ where
$\delta\geq 0$ and $\texttt{I}$ is an invariant
generated by bilinear contractions of the vectors $\texttt{V}$, $\texttt{U}$, $\texttt{W}$. These dressing factors
introduce invariants with higher and higher mass dimension. The defining relations (\ref{Vvec}), (\ref{Uvec}) and
(\ref{Wvec}) are too few to determine all the expansion coefficients in (\ref{delta-decomp}). Therefore, all one can do
is to express nine of the coefficients in terms of the rest. For instance, the coefficients in the linear sector can be
expressed in terms of those in the trilinear sector, leaving $\nu$'s, $\upsilon$'s and $\omega$'s undetermined, and
accordingly, all the invariants in (\ref{list}) can be expanded via (\ref{delta-decomp}) to determine the dynamics of
$\texttt{V}_{\alpha}$, $\texttt{U}_{\alpha}$, and $\texttt{W}_{\alpha}$. Nevertheless, as clearly suggested by
(\ref{delta-decomp}), the main effect of trilinear terms is to generate quartic and higher order interactions of
vectors. Putting emphasis on quadratic interactions, the trilinear terms can thus be left aside though they can be
straightforwardly included in the formulae below by processing the complete $\Delta^{\lambda}_{\alpha\beta}$ in
(\ref{delta-decomp}). Proceeding thus with linear terms in (\ref{delta-decomp}), one finds
\begin{eqnarray}
\label{coeffs}
a_v &=& c_v = a_u = b_u = b_w = c_w = -\frac{1}{18}\nonumber\\
b_v &=& c_u = a_w = \frac{5}{18}
\end{eqnarray}
for which $\Delta^{\lambda}_{\alpha\beta}$ gets decomposed linearly in terms of $\texttt{V}_{\alpha}$,
$\texttt{U}_{\alpha}$ and $\texttt{W}_{\alpha}$.

Given the decomposition in (\ref{delta-decomp}) of the tensorial connection, all the invariants in (\ref{list}) can be
expressed in terms of  $\texttt{V}_{\alpha}$, $\texttt{U}_{\alpha}$ and $\texttt{W}_{\alpha}$ to determine their
dynamics as vector fields hidden in the non-Riemannian geometry under consideration. To start with,
the curvature scalar $\mathbb{R}\left(g,\between\right)$, as follows from (\ref{decomp-R}), is composed of the GR part
$R\left(g,\Gamma\right)$ and the quasi curvature
scalar $g^{\alpha\beta} {\mathcal{R}}_{\alpha\beta}\left(\Delta\right) \equiv {\mathcal{R}}\left(g,\Delta\right)$. In
response to the linear part of the decomposition of $\Delta^{\lambda}_{\alpha\beta}$ in (\ref{delta-decomp}), the
latter takes the form
\begin{eqnarray}
\label{R-expand}
\mathcal{R}\left(g,\Delta\right)
&=& \nabla \cdot \left(\texttt{W} -\texttt{U}\right) +  \frac{1}{18}\Big( \texttt{V}\cdot \texttt{V} + \texttt{U}\cdot
\texttt{U} + \texttt{W}\cdot \texttt{W}\nonumber\\ &-& 4 \texttt{V}\cdot \texttt{U} - 4 \texttt{V}\cdot \texttt{W} + 14
\texttt{U}\cdot \texttt{W} \Big)
\end{eqnarray}
which shows that a term linear in $\mathbb{R}\left(g,\between\right)$ in the gravitational Lagrangian yields the
Einstein-Hilbert term $R\left(g,\Gamma\right)$ in GR plus a theory of three vector fields in which each vector develops
a `mass term' and  mixes with the others quadratically. The vectors do not acquire a kinetic term from
$\mathbb{R}\left(g,\between\right)$ since the first term at the right-hand side of (\ref{R-expand}), the divergence of
$\texttt{W}_{\alpha} - \texttt{U}_{\alpha}$, does not contribute to dynamics as it can be integrated out of the action
by using $\sqrt{-g} \nabla\cdot \left(\texttt{W} -\texttt{U}\right) = \partial_{\alpha} \left(\sqrt{-g}
\left(\texttt{W}^{\alpha} -\texttt{U}^{\alpha}\right)\right)$. One, however, notices that this term becomes important
in higher curvature terms like $\mathbb{R}^2\left(g,\between\right)$.

From (\ref{Rnew-acik}) it is already known that $\overline{\mathbb{R}}_{\alpha\beta}\left(\between\right)$ is the field
strength tensor of the vector field $\texttt{V}_{\alpha}$. Then the associated invariant in (\ref{list}) becomes
\begin{eqnarray}
\label{Rnew-expand}
\overline{\mathbb{R}} \bullet \overline{\mathbb{R}} =  \texttt{V}^{(-) \alpha\beta} \texttt{V}^{(-)}_{\alpha\beta}
\end{eqnarray}
which is nothing but the kinetic term of the Abelian vector $\texttt{V}_{\alpha}$.

Corresponding to the decomposition in (\ref{delta-decomp}), the torsion and non-metricity tensors take the explicit
form
\begin{eqnarray}
\label{torsion-acik2} \mathbb{S}^{\lambda}_{\alpha\beta} &=& \frac{1}{3} \left( \texttt{V}_{\alpha}
\delta^{\lambda}_{\beta} - \delta^{\lambda}_{\alpha} \texttt{V}_{\beta}\right)
- \frac{1}{3} \left( \texttt{U}_{\alpha} \delta^{\lambda}_{\beta} - \delta^{\lambda}_{\alpha}
\texttt{U}_{\beta}\right)\,,\\
\label{nonmet-acik2} \mathbb{Q}_{\lambda}^{\alpha\beta} &=& \frac{1}{9}\Big( 5 \texttt{V}_{\lambda} g^{\alpha\beta} -
\texttt{V}^{\alpha} \delta^{\beta}_{\lambda} - \delta^{\alpha}_{\lambda} \texttt{V}^{\beta}\nonumber\\
&-& \texttt{U}_{\lambda} g^{\alpha\beta} + 2 \texttt{U}^{\alpha} \delta^{\beta}_{\lambda} + 2 \delta^{\alpha}_{\lambda}
\texttt{U}^{\beta}\nonumber\\
&-& \texttt{W}_{\lambda} g^{\alpha\beta} + 2 \texttt{W}^{\alpha} \delta^{\beta}_{\lambda} + 2 \delta^{\alpha}_{\lambda}
\texttt{W}^{\beta}\Big)\,,
\end{eqnarray}
and thus, the related invariants in (\ref{list}) read as
\begin{eqnarray}
\label{sdots}
\mathbb{S}\bullet\mathbb{S} &=& 2 \left( \texttt{V}\cdot\texttt{V} + \texttt{U}\cdot\texttt{U} - 2
\texttt{V}\cdot\texttt{U} \right)\,,\\
\label{qdotq}
\mathbb{Q}\bullet\mathbb{Q} &=& \frac{2}{9} \Big( 22 \texttt{V}\cdot\texttt{V} + 7 \texttt{U}\cdot\texttt{U} + 7
\texttt{W}\cdot\texttt{W} + 20 \texttt{V}\cdot\texttt{U}\nonumber\\ &+& 20 \texttt{V}\cdot\texttt{W} + 14
\texttt{U}\cdot\texttt{W}\Big)\,,\\
\label{qdots} \mathbb{Q}\bullet\mathbb{S} &=& \frac{4}{3} \Big( 2
\texttt{V}\cdot\texttt{V} + \texttt{U}\cdot\texttt{U} - 3
\texttt{V}\cdot\texttt{U} - \texttt{V}\cdot\texttt{W}\nonumber\\
&+& \texttt{U}\cdot\texttt{W}\Big)\,.
\end{eqnarray}
This completes the decomposition of the quadratic invariants of the vector fields as generated by the curvature,
torsion and non-metricity tensors. It
is clear that these invariants provide a kinetic term only for $\texttt{V}_{\alpha}$; the other two vectors,
$\texttt{U}_{\alpha}$ and $\texttt{W}_{\alpha}$, acquire no kinetic term from any of the invariants in (\ref{list}).
Nevertheless,
a short glance at (\ref{torsion-acik2}) and (\ref{nonmet-acik2}) immediately reveals that the invariants formed by the
gradients of
curvature, torsion and non-metricity tensors can generate the requisite kinetic terms. Specifically, from
(\ref{torsion-acik2}) it is
found that
\begin{eqnarray}
\label{diver-torsion} \mathbb{D}_{\alpha \beta} =
\nabla^{\between}_{\lambda} \mathbb{S}^{\lambda}_{\alpha\beta}
\supset - \frac{1}{3} \texttt{V}^{(-)}_{\alpha\beta} + \frac{1}{3}
\texttt{U}^{(-)}_{\alpha\beta}
\end{eqnarray}
where the terms ${\cal{O}}\left(\Delta^2\right)$ are suppressed on the basis of unnecessity. The first term at the
right-hand
side is the field strength tensor of $\texttt{V}_{\alpha}$ as mentioned in (\ref{Rnew-acik}) and (\ref{fst-V}). The
second term is new in
that it is the field strength tensor of the $\texttt{U}_{\alpha}$ field. Therefore, divergence of torsion tensor
generates the requisite kinetic term for $\texttt{U}_{\alpha}$, and the associated invariant
\begin{eqnarray}
\label{kinetic-torsion}
 \mathbb{D} \bullet \mathbb{D} &\supset& \frac{1}{9}\Big( \texttt{V}^{(-) \alpha\beta} \texttt{V}^{(-)}_{\alpha\beta} +
 \texttt{U}^{(-) \alpha\beta} \texttt{U}^{(-)}_{\alpha\beta}\nonumber\\ &-& {2} \texttt{V}^{(-) \alpha\beta}
 \texttt{U}^{(-)}_{\alpha\beta}\Big)
\end{eqnarray}
encodes the kinetic terms of $\texttt{V}_{\alpha}$ and $\texttt{U}_{\alpha}$ as well as their kinetic mixing. One
notices that, not only the divergence operation (\ref{diver-torsion}) but also
\begin{eqnarray}
\label{torsion-extra}
g^{\rho \alpha} \nabla_{\rho}^{\between} \mathbb{S}^{\lambda}_{\alpha\beta} = -  g^{\rho \alpha}
\nabla_{\rho}^{\between} \mathbb{S}^{\lambda}_{\beta\alpha}
\end{eqnarray}
give contributions to the kinetic terms of vectors with similar structures as (\ref{kinetic-torsion}).

The candidate kinetic terms of $\texttt{V}_{\alpha}$ in (\ref{Rnew-expand}), and the kinetic term of
$\texttt{U}_{\alpha}$ in (\ref{kinetic-torsion})
are of the form expected of an $U(1)$ invariance. Of course, such an invariance is explicitly broken by the `mass
terms' generated by curvature,
torsion and non-metricity tensors. This is not the whole story, however. The kinetic terms generated by the derivatives
of the non-metricity
tensor in (\ref{nonmet-acik2}) also violate possible $U(1)$ invariance suggested by (\ref{Rnew-expand}) and
(\ref{kinetic-torsion}). To see this, one notes
that
\begin{eqnarray}
\label{kinetic-nonmet}
\mathbb{N}^{\alpha \beta} = g^{\rho\lambda} \nabla_{\rho}^{\between} \mathbb{Q}^{\alpha\beta}_{\lambda} &\supset&
\frac{1}{9}\Big( 5 \nabla\cdot\texttt{V} g^{\alpha\beta} - \texttt{V}^{(+) \alpha\beta}\nonumber\\
&-& \nabla\cdot\texttt{U} g^{\alpha\beta} + 2 \texttt{U}^{(+) \alpha\beta}\nonumber\\
&-& \nabla\cdot\texttt{W} g^{\alpha\beta} + 2 \texttt{W}^{(+) \alpha\beta}\Big)
\end{eqnarray}
where
\begin{eqnarray}
\texttt{V}^{(+)}_{\alpha\beta} \equiv \nabla_{\alpha} \texttt{V}_{\beta} + \nabla_{\beta}\texttt{V}_{\alpha}
\end{eqnarray}
is the symmetric counterpart of the anti-symmetric field strength tensor $\texttt{V}^{(-)}_{\alpha\beta}$ in
(\ref{fst-V}). This definition
holds also for the other vectors. Then the invariant generated by (\ref{kinetic-nonmet}) reads as
\begin{eqnarray}
\label{diver-nonmet}
\mathbb{N} \bullet \mathbb{N} \supset \frac{1}{162} \sum_{i,j=1}^{3} \texttt{A}^{(+)}_{i \alpha\beta} \texttt{K}_{i
j}^{\alpha\beta\mu\nu} \texttt{A}^{(+)}_{j\mu\nu}
\end{eqnarray}
where $\texttt{A}_i\in \left(\texttt{V}, \texttt{U}, \texttt{W} \right)$, and $\texttt{K}_{i j}^{\alpha\beta\mu\nu}$ is
the $(i,j)$-th entry of the matrix-valued tensor
\begin{eqnarray}
\label{pluscoeff}
\texttt{K}^{\alpha\beta\mu\nu} = \left(\begin{array}{lll}
 \texttt{K}^{\alpha\beta\mu\nu}_{1 1}  & \texttt{K}^{\alpha\beta\mu\nu}_{1 2}  &  \texttt{K}^{\alpha\beta\mu\nu}_{1 3} \\
\texttt{K}^{\alpha\beta\mu\nu}_{2 1} & \texttt{K}^{\alpha\beta\mu\nu}_{2 2} & \texttt{K}^{\alpha\beta\mu\nu}_{2 3}\\
\texttt{K}^{\alpha\beta\mu\nu}_{3 1} & \texttt{K}^{\alpha\beta\mu\nu}_{3 2} & \texttt{K}^{\alpha\beta\mu\nu}_{3 3} \end{array}\right)
\end{eqnarray}
where
\begin{eqnarray}
\texttt{K}^{\alpha\beta\mu\nu}_{1 1}&=&202 g^{\alpha\beta} g^{\mu \nu} + g^{\alpha\mu} g^{\beta \nu} + g^{\alpha\nu} g^{\beta\mu}\nonumber \\
\texttt{K}^{\alpha\beta\mu\nu}_{1 2}&=&\texttt{K}^{\alpha\beta\mu\nu}_{2 1}=
\texttt{K}^{\alpha\beta\mu\nu}_{1 3}= g^{\alpha\beta} g^{\mu\nu} - 2
g^{\alpha\mu} g^{\beta \nu} -2 g^{\alpha\nu} g^{\beta\mu}\nonumber \\
\texttt{K}^{\alpha\beta\mu\nu}_{2 2}&=&
\texttt{K}^{\alpha\beta\mu\nu}_{2 3}=\texttt{K}^{\alpha\beta\mu\nu}_{3 2}=\texttt{K}^{\alpha\beta\mu\nu}_{3 3}\nonumber \\&=&- 2
g^{\alpha\beta} g^{\mu\nu} + 4 g^{\alpha\mu} g^{\beta \nu} +4 g^{\alpha\nu} g^{\beta\mu}\nonumber
\end{eqnarray}
which describes the kinetic mixing among the three vector fields. As for the divergence of torsion in
(\ref{kinetic-torsion}), one notices that, not only the divergence operation (\ref{kinetic-nonmet}) but also
\begin{eqnarray}
\label{nonmet-extra}
\nabla_{\alpha}^{\between} \mathbb{Q}^{\alpha\beta}_{\lambda} = \nabla_{\alpha}^{\between} \mathbb{Q}^{\beta
\alpha}_{\lambda}
\end{eqnarray}
give contributions similar to that in (\ref{diver-nonmet}). In
addition to these, contraction of
\begin{eqnarray}
\label{cont} \nabla ^{\between} \mathbb{Q}\bullet
\nabla^{\between}\mathbb{S}&=& 0
\end{eqnarray}
due to symmetry conditions.

Having done with the decomposition of various invariants in terms of the vector fields $\texttt{V}$, $\texttt{U}$ and
$\texttt{W}$, we now turn to analysis of interactions in such a non-Riemannian setup. The most general action
functional describing `gravity' and `matter' is of the form
\begin{eqnarray}
\label{action-kapalý}
I = \int d^4x\ \sqrt{-g}\left\{ \mathfrak{L}\left(\mathbb{R}, \overline{\mathbb{R}}, \mathbb{S}, \mathbb{Q}\right) +
L_{m}\left(g, \between, \psi\right) - V_0\right\}
\end{eqnarray}
which contains action densities for geometric and material parts, respectively. $V_0$ stands for the vacuum energy
(containing the bare cosmological term fed by the geometrical sector), and  $\psi$ stands for matter and radiation
fields, collectively. Neither the geometrical $\mathfrak{L}$ nor the matter Lagrangian $L_m$ contains any constant
energy density; all such energy components are collected in $V_0$. The geometrical part reads explicitly as
\begin{eqnarray}
\label{action0} \mathfrak{L} &=& \frac{1}{2} M_{Pl}^2 \left(
\mathbb{R} + c_{S} \mathbb{S}\bullet\mathbb{S} + c_{Q}
\mathbb{Q}\bullet\mathbb{Q} + c_{QS}
\mathbb{Q}\bullet\mathbb{S}\right)\nonumber\\ &+&
c_S^{\prime} \nabla^{\between}\mathbb{S}\bullet \nabla^{\between}\mathbb{S} + c_Q^{\prime}
\nabla^{\between}\mathbb{Q}\bullet \nabla^{\between}\mathbb{Q} + c_{QS}^{\prime}
\nabla^{\between}\mathbb{Q}\bullet\nabla^{\between}\mathbb{S}\nonumber\\
&+& c_{R^2} \mathbb{R}^2 + c_{RR} \mathbb{R}\bullet\mathbb{R} +
c_{\overline{R}\overline{R}} \overline{\mathbb{R}}\bullet
\overline{\mathbb{R}} + {\mathcal{O}}\left(\frac{1}{M^2_{Pl}}\right)
\end{eqnarray}
where we have discarded terms ${\mathcal{O}}\left(1/M^2_{Pl}\right)$. Moreover, we have discarded higher-derivative
terms $\Box^{\between} \mathbb{R}$ and the like.
$c$'s are all dimensionless couplings.  The mass dimension-2 terms are naturally scaled by the fundamental scale of
gravity, $M_{Pl}$. One notices that
$\mathbb{R}^2$ and $\mathbb{R}\bullet\mathbb{R}$ contain higher-curvature terms $R(g,\Gamma)^2$ and
$R_{\alpha\beta}(\Gamma) R^{\alpha\beta}(\Gamma)$, respectively.  Indeed,
leaving aside the non-dynamical terms, one can show that
\begin{eqnarray}
\label{ricci-scalar}
\mathbb{R}^{2}\left(\between\right) \supset
R(g,\Gamma)^2+\Big((\nabla.\texttt{W})^{2}-2(\nabla.\texttt{W})(\nabla.\texttt{U})+(\nabla.\texttt{U})^{2}\Big)\nonumber\\~~~
\end{eqnarray}
and
\begin{eqnarray}
\label {ricci contraction}
\mathbb{R}\left(\between\right)\bullet\mathbb{R}\left(\between\right) &\supset& R^2(g,\Gamma)+ R_{\mu \nu}(g,\Gamma)R^{\mu\nu}(g,\Gamma)\nonumber \\
&+&\frac{1}{648}\Big(-4(\nabla.\texttt{V})^{2}+162(\nabla.\texttt{U})^{2}+167(\nabla.\texttt{W})^{2}\nonumber \\
&-&330(\nabla.\texttt{U})(\nabla.\texttt{W})-6(\nabla.\texttt{V})(\nabla.\texttt{U})\nonumber \\
&+&4(\nabla.\texttt{V})(\nabla.\texttt{W})+16\nabla_{\mu}\texttt{V}_{\nu}\nabla^{\nu}\texttt{V}^{\mu}\nonumber \\&+&24\nabla_{\mu}\texttt{V}_{\nu}\nabla^{\nu}\texttt{U}^{\mu}
+9\nabla_{\mu}\texttt{U}_{\nu}\nabla^{\mu}\texttt{U}^{\nu}\nonumber \\&+&10\nabla_{\mu}\texttt{V}_{\nu}\texttt{V}^{(-)\mu\nu}-18\nabla_{\mu}\texttt{V}_{\nu}\texttt{U}^{(-)\mu\nu}\nonumber \\
&-&8\nabla_{\mu}\texttt{V}_{\nu}\texttt{W}^{(+)\mu\nu}+8\nabla_{\mu}
\texttt{U}_{\nu}\texttt{U}^{(-)\mu\nu}\nonumber\\
&-&6\nabla_{\mu}\texttt{U}_{\nu}\texttt{W}^{(+)\mu\nu}+2\nabla_{\mu}\texttt{W}_{\nu}\texttt{W}^{(+)\mu\nu}\Big)
\end{eqnarray}
wherein the GR-related parts are seen to involve higher-derivative interactions. In this sense, the GR-part (the terms
$R^2(g,\Gamma)$ and $R_{\mu \nu}(g,\Gamma)R^{\mu\nu}(g,\Gamma)$ ) brings forth ghosts. Clearly, these terms must be
absent ($c_{R^2}$ and $c_{RR}$ must vanish) if such ghosty contributions in GR are to be avoided. The remaining terms, after using
their decompositions in terms of the vector fields $\texttt{V}$, $\texttt{U}$ and $\texttt{W}$, give rise to the
action
\begin{widetext}
\begin{eqnarray}
\label{action-acik}
I &=& \int d^4x\ \sqrt{-g} \left\{ \frac{1}{2} M_{Pl}^2 R + L_{m}\left(g,\between,\psi\right) - V_0\right\} \nonumber\\
&+& \int d^4x\ \sqrt{-g} \Bigg\{ c_{VV} \texttt{V}^{(-)\alpha\beta} \texttt{V}^{(-)}_{\alpha\beta} + c_{UU}
\texttt{U}^{(-)\alpha\beta} \texttt{U}^{(-)}_{\alpha\beta} + c_{VU} \texttt{V}^{(-)\alpha\beta}
\texttt{U}^{(-)}_{\alpha\beta}\nonumber\\
&+& \texttt{V}^{(+)}_{\alpha\beta} \texttt{k}_{VV}^{\alpha\beta\mu\nu}\texttt{V}^{(+)}_{\mu\nu} +
\texttt{U}^{(+)}_{\alpha\beta} \texttt{k}_{UU}^{\alpha\beta\mu\nu}\texttt{U}^{(+)}_{\mu\nu} +
\texttt{W}^{(+)}_{\alpha\beta} \texttt{k}_{WW}^{\alpha\beta\mu\nu}\texttt{W}^{(+)}_{\mu\nu} +
\texttt{V}^{(+)}_{\alpha\beta} \texttt{k}_{VU}^{\alpha\beta\mu\nu}\texttt{U}^{(+)}_{\mu\nu} +
\texttt{V}^{(+)}_{\alpha\beta} \texttt{k}_{VW}^{\alpha\beta\mu\nu}\texttt{W}^{(+)}_{\mu\nu} +
\texttt{U}^{(+)}_{\alpha\beta} \texttt{k}_{UW}^{\alpha\beta\mu\nu}\texttt{W}^{(+)}_{\mu\nu} \nonumber\\
&+&  M_{Pl}^2 \Big( \frac{1}{2} a_{VV} \texttt{V}^{\alpha} \texttt{V}_{\alpha} + \frac{1}{2} a_{UU}\texttt{U}^{\alpha}
\texttt{U}_{\alpha} + \frac{1}{2} a_{WW}\texttt{W}^{\alpha} \texttt{W}_{\alpha} + a_{VU} \texttt{V}^{\alpha}
\texttt{U}_{\alpha} + a_{VW} \texttt{V}^{\alpha} \texttt{W}_{\alpha} + a_{UW} \texttt{U}^{\alpha} \texttt{W}_{\alpha}
\Big) \Bigg\}
\end{eqnarray}
\end{widetext}
where the first integral at the right-hand side is precisely the Einstein-Hilbert action in GR (plus the contribution
of matter and radiation), and the second integral pertains to a theory of three vector fields in a spacetime with
metric $g_{\alpha\beta}$. The Einstein-Hilbert action above would receive contributions from
higher-curvature (and thus typically ghosty) terms had we kept $\mathbb{R}^2$ and $\mathbb{R}\bullet\mathbb{R}$ terms
in (\ref{action0}).

In essence, under the decomposition in (\ref{delta-decomp}), the non-Riemannian gravitational theory in (\ref{action0})
reduces
to a tensor-vector theory of the type in (\ref{action-acik}) (leaving aside the matter sector
$L_{m}\left(g,\between,\psi\right)$). One notices that the general connection $\between^{\lambda}_{\alpha\beta}$ can directly couple to matter fields as encoded in the matter Lagrangian. According to types of the matter fields, these couplings give rise to additional
structures (like hyper-momentum) which involve torsion and non-metricity. In \cite{dynamics1,dynamics2}, various effects of the general connection on the matter sector are analysed in detail.
 The vector part of the action is written in a rather generic form by admitting that various terms listed above plus
 similar
  ones coming, for example, from (\ref{torsion-extra}) and
(\ref{nonmet-extra}) give rise to, at the quadratic level, the structures in (\ref{action-acik}) with dimensionless
coefficients
$c_{VV},\dots, a_{UW}$. These coefficients can be expressed as linear combinations of the coefficients weighing
individual contributions.

The tensor-vector theory in (\ref{action-acik}) has been obtained for a general setup involving curvature, torsion and
non-metricity tensors
exhaustively. The theory is GR plus a theory of three vectors $\texttt{V}$, $\texttt{U}$ and $\texttt{W}$. Any
constraint or selection rule
imposed on the non-Riemannian geometry results in a more restricted theory. It could thus be useful to discuss certain
aspects of (\ref{action-acik})
here:
\begin{itemize}
\item Theory consists of three vector fields $\texttt{V}$, $\texttt{U}$ and $\texttt{W}$. The vector action
    contains two types of kinetic
terms: ones with $\texttt{X}^{(-)}_{\alpha\beta}$ and those with $\texttt{X}^{(+)}_{\alpha\beta}$. The $\texttt{V}$
and  $\texttt{U}$ possess both types of kinetic terms while $\texttt{W}$ possesses only the second type {\it i. e.}
$\texttt{W}^{(+)}_{\alpha\beta}$. The $\texttt{X}^{(-)}_{\alpha\beta}$ and hence the corresponding kinetic terms
obviously possess an Abelian invariance. However, there is no such invariance for the kinetic terms involving
$\texttt{X}^{(+)}_{\alpha\beta}$.
Therefore, the vector fields contained in (\ref{action-acik}) are not associated with a gauge theory; they are not
vectors originating from need to realize a local U(1) invariance.

The coefficients $c_{VV},\dots, k_{UV}^{\alpha\beta\mu\nu}$, which seem being left arbitrary, can actually be fixed
in terms of the coefficients of
individual terms in (\ref{action0}) which contribute to that particular structure. The kinetic terms, both
$\texttt{X}^{(-)}_{\alpha\beta}$ and $\texttt{X}^{(+)}_{\alpha\beta}$ type, receive contributions from various
structures,  as addressed before. In particular, contributions of the alternative structures
given in (\ref{torsion-extra}) and (\ref{nonmet-extra}) must also be included in forming the vector action in
(\ref{action-acik}).

\item A highly crucial aspect concerns the signs of the coefficients $c_{VV},\dots, k_{UV}^{\alpha\beta\mu\nu}$ in
    the kinetic part of the vector action. The kinetic
terms of $\texttt{V}$, $\texttt{U}$ and $\texttt{W}$ must have the correct sign required of a ghost-free theory.
Indeed, any sign-flip in the kinetic terms causes vector ghosts to show up in the spectrum. The various
coefficients in (\ref{action0}) must comply with this requirement.

\item The vectors exhibit not only the kinetic mixings  $\texttt{X}^{(-)}_{\alpha\beta} \texttt{Y}^{(-)
    \alpha\beta}$ and $\texttt{X}^{(+)}_{\alpha\beta} \texttt{Y}^{(+) \alpha\beta}$ but also mass mixings of the
    form $\texttt{X}^{\alpha} \texttt{Y}_{\alpha}$, as shown in the last line of the vector action. Their
    masses and mixings are proportional to $M_{Pl}$ with respective coefficients $a_{VV},\dots,a_{UW}$. In
    $\left\{\texttt{V}, \texttt{U},  \texttt{W}\right\}$
basis their mass-squared matrix reads as
\begin{eqnarray}
\label{mass-mat}
\frac{1}{2} M_{Pl}^2 \left(\begin{array}{ccc} a_{VV} & a_{VU} & a_{VW}\\  a_{VU} & a_{UU} & a_{UW}\\ a_{VW} &
a_{UW} & a_{WW} \end{array} \right)
\end{eqnarray}
each entry of which can be extracted from (\ref{action0}) as
\begin{eqnarray}
\label{relatex}
a_{VV} &=& \frac{1}{18} + 2 c_S + \frac{44}{9} c_{Q} + \frac{8}{3} c_{QS}\,,\nonumber\\
a_{UU} &=& c_{WW} +  2 c_S + \frac{4}{3} c_{QS}\,,\nonumber\\
a_{WW} &=& \frac{1}{18} + \frac{14}{9} c_{Q}\,,\nonumber\\
a_{VU} &=& -\frac{1}{9} -2 c_S + \frac{20}{9} c_{Q} - 2 c_{QS}\,,\nonumber\\
a_{VW} &=& -\frac{1}{9} + \frac{20}{9} c_{Q} - \frac{2}{3} c_{QS}\,,\nonumber\\
a_{UW} &=& \frac{7}{18} + \frac{14}{9} c_{Q} + \frac{2}{3} c_{QS}\,.
\end{eqnarray}
It is the eigenvalues of (\ref{mass-mat}) that determine the light and heavy vector spectrum in the theory. For having a stable
theory free from tachyons, the eigenvalues of (\ref{mass-mat}) must each be positive semi-definite. This puts stringent
constraints on the elements $a_{VV},\cdots, a_{UW}$ (See Appendix B for further details.). If
off-diagonal entries are small {\it i. e.} if $c_S$, $c_Q$ and $c_{QS}$ are chosen appropriately then all three
vector bosons weigh $M_{Pl}/3\sqrt{2}$. Alternatively, if the mixings are sizeable, or equivalently, if all entries
of (\ref{mass-mat}) are of similar size then there will exist two light and one heavy vectors in the spectrum.
Depending on the hierarchy of the couplings, there could exist just one light state instead of two \cite{DAD}. In
any case, it is with the hierarchy of the couplings that the vector boson spectrum can exhibit different
hierarchies. Needless to say, the intra-hierarchy of the mass matrix entries $a_{VV},\dots,a_{UW}$ is determined by
the couplings $c_S$, $c_Q$ and $c_{QS}$ via the relations (\ref{relatex}).

Actually, having the vector fields with masses around $M_{Pl}$ should come by no surprise; the underlying theory
(\ref{action0}) is a pure gravity of non-Riemannian structure, and the mass scale in the theory is automatically
fixed by the fundamental scale of gravity $M_{Pl}$. However, the statement `a Planckian-mass vector field' depends
crucially on what we mean by the vector field: Is it fundamental or is it a composite structure? We will discuss
answers and consequences of these questions in the sequel.

\item As is obvious from the general procedure, reduction of the non-Riemannian gravity gives rise to GR plus
extra degrees of freedom represented by the vector fields in (\ref{action-acik}). These extra degrees of freedom
can have astrophysical and cosmological implications, and can give rise to observable phenomena at high-energy
particle colliders. These fields may form an invisible sector which couples to known matter via Higgs or
vector boson portals. We shall discuss some of their cosmological effects in the next section.

\item The framework we have reached in (\ref{action-acik}) is a rather general one in that we have imposed no
    condition on metric, connection and any other geometro-dynamical quantity. Imposition of certain selection
    rules, though seems to cause loss of generality, does actually prove highly useful for
    extracting information about behavior of the system in certain reasonable situations. Here we shall discuss two
    such limiting cases:
    \begin{itemize}
    \item {\bf Symmetric Connection:} We first discuss the possibility of symmetric connection {\it i. e.}
        $\between^{\lambda}_{\alpha \beta} = \between^{\lambda}_{\beta \alpha}$. The prime implication of this
        selection rule is that the torsion tensor identically vanishes, $\mathbb{S}^{\lambda}_{\alpha\beta} =
        0$. This statement is equivalent to imposing
        \begin{eqnarray}
        \label{symm}
        \texttt{V}_{\alpha} = \texttt{U}_{\alpha},
        \end{eqnarray}
        as is manifestly suggested by the decomposition of $\Delta^{\lambda}_{\alpha\beta}$ in
        (\ref{delta-decomp}). This constraint is seen to nullify the invariants $\mathbb{S}\cdot\mathbb{S}$ and
        $\mathbb{S}\cdot\mathbb{Q}$, in agreement with vanishing torsion. This particular relation between
        $\texttt{V}$ and $\texttt{U}$ reduces the vector action in (\ref{action-acik}) into a theory of two
        vectors: the $\texttt{V}$ and $\texttt{W}$. The structure remains similar to that in
        (\ref{action-acik}) yet various terms containing $\texttt{V}$ and $\texttt{U}$ merge together to give
        more compact relations.

    \item {\bf Antisymmetric Tensorial Connection:} This time we consider  the relation
        $\Delta^{\lambda}_{\alpha\beta} = - \Delta^{\lambda}_{\beta\alpha}$ for the tensorial connection not
        for $\between^{\lambda}_{\alpha\beta}$. Actually, since $\Gamma^{\lambda}_{\alpha\beta}$ is symmetric
        the connection $\between^{\lambda}_{\alpha\beta}$ possesses no obvious symmetry under the exchange of
        $\alpha$ and $\beta$. The prime implication of the anti-symmetric $\Delta$ is that the geodesics of
        test bodies remain as in the GR. This, however, does not mean that one can eliminate the non-Riemannian
        effects. The reason is that the geodesic deviation, which involves the Riemann tensor
        $\mathbb{R}^{\alpha}_{\mu\beta\nu}$, directly feels the non-GR components of the curvature tensor. In
        the language of the expansion (\ref{delta-decomp}), anti-symmetric $\Delta^{\lambda}_{\alpha\beta}$
        gives
        \begin{eqnarray}
        \label{antisymm}
        \texttt{V}_{\alpha} = - \texttt{U}_{\alpha}\,\,\;\mbox{and}\;\; \texttt{W}_{\alpha} = 0
        \end{eqnarray}
        which reduces thus the vector action in (\ref{action-acik}) to theory of a single vector field
        $\texttt{V}$.
    \end{itemize}
\end{itemize}

Here we have highlighted certain salient features of the Tensor-Vector theory of (\ref{action-acik}) in regard to
various structures and limiting cases the vector part can take.

\section{APPLICATIONS TO COSMOLOGY}

Up to now, we have constructed a general action which consists of
all possible vector and tensor fields. In addition to this, we have
given two limiting cases as symmetric and antisymmetric tensorial
connection. In next two subsections, by using antisymmetric
tensorial connection limit and some constraints, we obtain two
well-known actions which are defined in modified gravity theories
These are TeVeS gravity and Vector Inflation.

\subsection{TeVeS Gravity}

In spite of its great success in describing the solar system, General Relativity (GR) fails to account for dynamics
at galactic scales without postulating a large amount of cold dark matter (CDM) -- non-baryonic, non-relativistic,
electrically neutral, weakly interacting particles of weak-scale masses \cite{DM}. The asymptotic flatness of the
galaxy
rotation curves, which occurs towards galaxy outskirts involving extremely small accelerations, manifestly disagrees
with predictions of the GR unless the galactic region is populated by non-shining, and hence, astrophysically
unobservable
CDM.

Apart from this, there are problems with structure formation: with the baryonic matter alone, the large-scale
structure as we observe it would not have been formed yet if gravity is described by GR. Indeed, GR demands
large amounts of `dark components' ($23\%$ `dark matter' for structure formation and $73\%$ `dark energy' for
late-time inflation) to be able to account for the mounting cosmo-physical precision data (coming from observations
on microwave background \cite{observe1}, large scale structure \cite{observe2}, and supernovae \cite{observe3}).
However,
the way these dark components enter into gravitational field equations does not involve their origins and nature;
they are treated as `fluids' with right density and equation of state. Nevertheless, the positron excess reported by
recent observations \cite{cosmic} on cosmic rays, if interpreted to come from decays or annihilations of dark matter,
can be taken as indirect signals of dark matter (though there are alternative arguments in favor of
astrophysical sources \cite{cosmic2} of positron excess).

This 'dark paradigm' necessitated by GR can in fact be evaded if an alternative description of
Nature takes over at extremely small accelerations and curvatures. This is what has been postulated
by Milgrom \cite{milgrom}, who replaced Newton's second law of motion with
\begin{eqnarray}
\label{milgrom}
\mu\left(\frac{\left|\vec{a}\right|}{a_0}\right) \vec{a} = - \vec{\nabla} \Phi_N
\end{eqnarray}
where $\Phi_N$ is the gravitational potential, $\mu(x) \leadsto 1 (x)$ for $x\gg 1 (x \ll 1)$, and
$a_0 \simeq 10^{-10}\ m/s^2$ is an acceleration scale appropriate for galaxy outskirts \cite{Mond-f}. This proposal,
despite its empirical success, had to wait for the relativistic generalizations of \cite{sanders,bekenstein-mond}
to become a complete, alternative theory of gravitational interactions (see also the review \cite{review}). The
relativistic generalization, dubbed as
tensor-vector-scalar (TeVeS) theory of gravity, involves the {\it geometrical} fields $V_{\mu}$ and $\phi$ in addition
to the metric tensor $g_{\mu \nu}$ such that, while the matter sector involves $g_{\mu \nu}$ only, the gravitational
sector involves
\begin{eqnarray}
\label{gen-metric}
\widetilde{g}_{\mu \nu} = e^{2 \phi} g_{\mu \nu} - 2 \sinh (2 \phi) A_{\mu} A_{\nu}
\end{eqnarray}
whose action can be generalized to incorporate aether effects \cite{aether}, too. Various astrophysical and
cosmological phenomena exhibit observable signatures of TeVeS \cite{bekenstein-mond2,seljak}.

TeVeS is essentially a bi-metrical gravitational theory where matter and gravity are distinguished by
the metric fields they operate with. It is thus natural to expect a reformulation in bi-metrical language
\cite{bi-theory}
wherein certain interactions and properties follow deductively.

The material produced in the last section is general and detailed enough to have a re-look at the TeVeS gravity. In
this section we will argue that TeVeS type extended gravity theories
do naturally follow from the non-Riemannian theories of the form (\ref{action0}) under the decomposition
(\ref{delta-decomp}).

\begin{itemize}
\item To establish contact with TeVeS gravity, it is necessary to discuss first the function $\mu$  defined in
    (\ref{milgrom}). In relativistic
formulation,  $\mu$ is a non-dynamical field in the action. Variation of the action with respect to $\mu$ fixes
`gradient' of
its potential $d V(\mu)/d\mu$ in terms of the remaining terms in which $\mu$ appears at least linearly. Basically,
$\mu$ must
multiply the kinetic term of (scalars or vectors) so that its force $d V(\mu)/d\mu$ is fixed in terms of the field
gradient-squareds
(actually the kinetic terms of the fields) in accord with the requirements of the MOND. In summary, the MOND
relation (\ref{milgrom})
for $\mu$ arises from the equation of motion for $\mu$ (to be solved via $d V(\mu)/ d\mu$ in terms of the kinetic
terms of the
fields in the spectrum). The relativistic theory of \cite{bekenstein-mond} requires that $\mu$ should be
non-dynamical, that is, it
should have no kinetic term. Therefore, the Lagrangian of $\mu$ can be directly constructed from couplings in the
action (\ref{action-acik}).
We do this as follows:
\begin{itemize}
\item First, we postulate that  the vacuum energy density $V_0$ in (\ref{action-kapalý}) and (\ref{action-acik}) can
    actually be
decomposed as
\begin{eqnarray}
V_0 = V\left(\mu\right) + \Delta V
\end{eqnarray}
where $\Delta V$ is a constant additive energy density while $V$ varies with $\mu$. At this stage $\mu$ is a
hypothetical parameter
having no solid physical basis.

\item We further postulate that the coefficients $c_{VV},\dots, k_{UV}^{\alpha\beta\mu\nu}$ weighing the
    individual kinetic terms
in the vector part of the action (\ref{action-acik}) do actually depend on the parameter $\mu$ at least in a
linear fashion. In
fact, it is not necessary to make all these constants vary with $\mu$; the $\mu$ dependence of one single
parameter suffices.
\end{itemize}

\item Under these instructed changes for `creating' or `explicating' the non-dynamical field $\mu$, the action
    (\ref{action-acik})
becomes essentially the Tensor-Vector Theory of \cite{aether}. This theory is obtained by eliminating  the scalar
field through
the constraints on the bimetrical theory of \cite{bekenstein-mond}, and is shown to be a viable replacement for
cold dark matter.
As an aether theory, it works as good as the model in \cite{bekenstein-mond} as far as the MOND-change of gravity
is concerned.
The main distinction between the theory obtained here and that of \cite{aether} is that  the model here consists of
three vectors
in the most general case. If one specializes to cases like (\ref{symm}) or (\ref{antisymm}), however, the model
obtained here gets closer to
the aether theory of \cite{aether}, which is shown therein to be an alternative to the cold dark matter.

Consequently, the Tensor-Vector theory in (\ref{action-acik}) provides a general enough framework (in terms of
parameters and
number of vector fields) for generating  the TeVeS gravity of \cite{milgrom, sanders, bekenstein-mond, review}
through the
analyses in \cite{aether}. It should be kept in mind that, the TeVeS gravity of \cite{sanders,bekenstein-mond} is
based on a bimetrical
theory where the geometrical sector proceeds with metric involving a scalar field, vector field and the metric
field used by the matter Lagrangian. The theory in the present work, however, provides a compact approach to TeVeS
gravity via
the decomposition of the tensorial connection in (\ref{delta-decomp}).

\item At this point, one may wonder why we are dealing with the Tensor-Vector theory of \cite{aether} instead of
    the true TeVeS gravity
of \cite{sanders,bekenstein-mond,review}. Actually, as we will shown below, the action (\ref{action-acik})
naturally contains the true
TeVeS gravity. To this end, the right question to ask concerns the vector fields themselves: Are they fundamental
vector fields or composites of some other fields?
They each could be of either nature. Whatever their structure, however, they must be true vectors on the spacetime
manifold such that
their vector property must not depend on the connections $\between^{\lambda}_{\alpha\beta}$ or
$\Gamma^{\lambda}_{\alpha\beta}$ or
$\Delta^{\lambda}_{\alpha\beta}$. The reason is that the vectors themselves are just parameterizing  the connection
via (\ref{delta-decomp}), and hence,
their independence from the connection is required by the logical consistency of the construction. This constraint
prohibits all structures but
\begin{eqnarray}
\label{V-decomp}
\texttt{V}_{\alpha} = a_{1} V_{\alpha} + \frac{a_0}{M_{Pl}} \partial_{\alpha}\phi
\end{eqnarray}
where $V_{\alpha}$ is a fundamental vector, and $\phi$ is a fundamental scalar field. The
vector property of $V_{\alpha}$ is obvious. Why the $\phi$-dependent part is a vector
is guaranteed by the fact that $\nabla_{\alpha}^{(any\ connection)}\phi = \partial_{\alpha} \phi$,
and hence, it is a vector on the manifold independent of the connection; may it be
$\between^{\lambda}_{\alpha\beta}$ or
$\Gamma^{\lambda}_{\alpha\beta}$ or some other structure. Obviously, if $\phi$ is to be a
new degree of freedom (not a scalar formed form $V_{\alpha}$ itself) then it is necessary to
reduce the degrees of freedom contained in $V_{\alpha}$ by one unit. Any `gauge constraint'
such as $\nabla \cdot V = 0$ proves sufficient for this purpose. Under these conditions,
the expansion (\ref{V-decomp}) operates on each of the vectors $\texttt{V}$,
$\texttt{U}$ and $\texttt{W}$ with their respective scalar fields.

It is obvious that replacement of (\ref{V-decomp}) and similar relations for $\texttt{U}_{\alpha}$ and
$\texttt{W}_{\alpha}$
into the vector action in (\ref{action-acik}) will yield a general tensor-vector-scalar theory of gravity. The
main
difference from \cite{sanders,bekenstein-mond,review} will be the number of vectors and scalars in the theory. The
difference
will be the dependence of the action on the scalars: Only the gradients of scalars are involved. The scalars
themselves do not
enter the action. Nevertheless, one arrives at a tensor-vector-scalar theory of gravity, and the theory is
parametrically and
dynamically wide enough to cover the standard TeVeS gravity.

\item As a concrete case study, here we shall discuss the reduced theory after imposing the condition
    (\ref{antisymm}). The
action (\ref{action-acik}) reduces to
\begin{eqnarray}
\label{action-acik2}
I &=& \int d^4x\ \sqrt{-g} \Bigg\{ \frac{1}{2} M_{Pl}^2 R + L_{m}\left(g,\between,\psi\right) - V_0 \nonumber\\
&+& \overline{c}_{VV} \texttt{V}^{(-)\alpha\beta} \texttt{V}^{(-)}_{\alpha\beta}
+ \texttt{V}^{(+)}_{\alpha\beta}
\overline{\texttt{k}}_{VV}^{\alpha\beta\mu\nu}\texttt{V}^{(+)}_{\mu\nu}\nonumber\\
&+&  \frac{1}{2} M_{Pl}^2  \overline{a}_{VV} \texttt{V}^{\alpha} \texttt{V}_{\alpha} \Bigg\}
\end{eqnarray}
where the terms involving $\texttt{V}$ and $\texttt{U}$ in (\ref{action-acik}) combine to form the
over-lined coefficients in here. The terms involving $\texttt{W}$ in (\ref{action-acik}) are all nullified
in accord with (\ref{antisymm}). For instance, one directly finds
\begin{eqnarray}
\overline{a}_{VV} = \frac{1}{3} + 8 c_S + 2 c_Q + 8 c_{QS}
\end{eqnarray}
form (\ref{relatex}). The reduced theory in (\ref{action-acik2}) is
precisely the one in \cite{aether} except for the absence of
quartic-in-$\texttt{V}$ terms. The couplings in and dynamics of the
two theories can be matched via the terms involved in two cases.
This situation becomes especially clear after using
$\texttt{V}^{\alpha}\texttt{V}_{\alpha} = -1$ in the tensor-vector
theory of \cite{aether}.

Now, it is time to analyze (\ref{action-acik2}) under the decomposition (\ref{V-decomp}). One finds
\begin{eqnarray}
\label{action-acik3}
I &=& \int d^4x\ \sqrt{-g} \Bigg\{ \frac{1}{2} M_{Pl}^2 R + L_{m}\left(g,\between,\psi\right) - V_0 \nonumber\\
&+& a_1^2 \overline{c}_{VV} {V}^{(-)\alpha\beta} {V}^{(-)}_{\alpha\beta}
+ a_1^2 {V}^{(+)}_{\alpha\beta} \overline{\texttt{k}}_{VV}^{\alpha\beta\mu\nu} {V}^{(+)}_{\mu\nu}\nonumber\\
&+&  \frac{1}{2} M_{Pl}^2  a_1^2 \overline{a}_{VV} {V}^{\alpha} {V}_{\alpha}
+  M_{Pl}  a_1 a_0 \overline{a}_{VV} {V}^{\alpha} \partial_{\alpha}\phi\nonumber\\
&+& a_0^2 \overline{a}_{VV} {\partial}^{\alpha}\phi \partial_{\alpha}\phi + {\cal{O}}\left(\frac{1}{M_{Pl}}\right)
\Bigg\}
\end{eqnarray}
from which it is seen that setting $V_0 \equiv V(\mu) + \Delta V$
and $a_0 = \bar{a}_0 \mu$ essentially suffices to reproduce the
results of TeVeS gravity \cite{bekenstein-mond,review}. Setting
$V^{\alpha}V_{\alpha} =-1$ as a constraint on the vector field, the
mass term of $V^{\alpha}$ in (\ref{action-acik3}) just adds up to
the vacuum energy $V_0$.
\end{itemize}

Before closing this section we comment on MOND. The MOND theory (or its relativistic realization TeVeS) has been put forth as an alternative to the Dark Matter paradigm. As for any model, there are phenomena for which TeVeS cannot give a satisfactory explanation. Indeed, while it can explain flat rotation curves with no need to Dark Matter, it has phenomenological shortcomings related to explanations of the other DM evidences such as Bullet Cluster. Nevertheless, like the Dark Matter paradigm all these models are under theoretical and experimental investigation, and one can find better
realizations in terms of various constraints. The non-Riemannian origin we discuss is not special to TeVeS or any other specific modeling; it holds
in general and its parameter space can be constrained by astrophysical observations or collider experiments.

\subsection{Vector Inflation}

According to the standart big bang cosmology, which is defined by
using Friedmann-Robertson-Walker (FRW) metric, universe is
homogeneous and isotropic on large scales \cite{Liddle-Lyth}. In addition
to this, observations of Hubble in redshifts of galaxies shows that
universe is expanding. To understand dynamical properties of
expansion, the solutions of Einstein equation for FRW metric are
required. Combination of these solutions is given by
\begin{eqnarray}
\label {acceleration}
\frac{\ddot{a}}{a}=-\frac{4\pi}{3M_{pl}^{2}}(\rho+3p)
\end{eqnarray}
as $\rho$ implies energy density and $p$ is pressure and $a$ is
scale factor. In the ligth of equation (\ref{acceleration}) one can
think that universe expands by decelerating in case of
$(\rho+3p)>0$. However, this deceleration doesn't solve some problem
of standart big bang cosmology such as flatness, horizon and so on.
To solve these problems, accelerated expansion of universe in early
stage is treated instead of decelerated one i.e $(\rho+3p)<0$ and
this type of expansion is called "inflation". Inflation is generally
driven by scalar fields to prevent anisotropy occured in higher spin
fields \cite {initial}. However, scalar inflation models have
fine-tuning problem and also scalar bosons which is base of these
models aren't observed by experiments\cite{weyl's}. Therefore,
vector inflation model is condsidered instead of scalar inflation
model. \cite{vecinf1,vect-inf} Also p-forms inflation model is also considered in
literature\cite{p-nflation}.

 Vector inflation was firstly proposed in \cite{vecinf1} by using spacelike vector fields.
In Ford's paper vector fields gave anisotropic solution of
inflation. So, instead of spacelike vector fields, it was shown that
timelike vector fields under some constraints of vector field
potential give rise to desired inflationary expansion
\cite{vecinf2}. The other problems vector fields have can be solved
by using a triplet of mutually orthogonal vector fields and non-
minimally coupling.

In this section, we show that after a regularization ,the action
(\ref{action-acik}) obtained by using the anti-symmetric connection
constraint give the same action in \cite{vecinf2} which is  most
general action of vector inflation theory.

Combining the abelian and non-abelian part of vector field and
defining new dimensionless coefficients lead to the action (leaving aside the matter sector):
\begin{widetext}
\begin{eqnarray}
\label{vector-inf}
I&=&\int d^4x\ \sqrt{-g} \Bigg\{ \frac{1}{2}
M_{Pl}^2 R+\frac{1}{2} \mathcal{\kappa}^{\alpha \beta \mu
\nu}\nabla_{\alpha}{V}_{\beta}\nabla_{\mu}{V}_{\nu}+V(\xi)\Bigg\}\nonumber\\~~~~~~
\end{eqnarray}
\end{widetext}
where
\begin{eqnarray}
V(\xi)= \frac{1}{2}M_{Pl}^2
\overline{a}_{VV}{V}^{\alpha}{V}_{\alpha}
\end{eqnarray}
and
\begin{eqnarray}
\label{kappaa}
 \mathcal{\kappa}^{\alpha \beta \mu \nu}= \kappa_{1}g^{\alpha \beta}g^{\mu \nu}+\kappa_{2}g^{\alpha \mu}g^{\beta \nu}+
 \kappa_{3}g^{\alpha \nu}g^{\beta \mu}
 \end{eqnarray}
 $\xi= {V}^{\alpha}{V}_{\alpha}$, and $\kappa_{1}, \kappa_{2},\kappa_{3}$ are random coefficients coming
 from general action.
 \begin{eqnarray}
\kappa_{1}&=& \frac{44}{18}c_{Q}^{'},\nonumber\\
\kappa_{2}&=& \frac{8c_{s}^{'}+2c_{Q}^{'}}{18},\nonumber\\
\kappa_{3}&=& \frac{2c_{Q}^{'}-8c_{s}^{'}}{18}
\end{eqnarray}

Assigning suitable values (by excluding ones leading to linear
instabilities or negative-energy ghosts) to these coefficients reproduce the same
results with the action of general vector inflation in
\cite{vecinf2}.

\section{Conclusion}

Metric-affine gravity generalizes the GR by accommodating an affine
connection that extends the Levi-Civita connection. The tensorial
part of the connection, under general conditions, can be decomposed
into three independent vector fields (and a fundamental rank (1,2) tensor
field, if any) which can be fundamental fields or gradients of some scalar fields.
By this way the vector, scalar and tensor fields come into play when
the metric-affine action is decomposed accordingly. The resulting
theory is rather general. By imposing judicious constraints, theory can be
reduced to more familiar ones like TeVeS gravity, vector inflation or
aether-like models, in general. In the text we have given a detailed
discussion of the TeVeS gravity and vector inflation.

From this work, one concludes that metric-affine gravity is rich enough to
supply various vector and scalar fields needed in cosmological phenomena. Analyses
of various effects may lead to a standard model of metric-affine gravity.

{\bf Acknowledgements}\\
 C.~N.~K. thanks Glenn Starkman and Fred Adams for their comments, criticisms and suggestions. A.~A. thanks Selin Soysal for discussions. We would like to thank V. Vitagliano and very conscientious referees for their useful comments and suggestions.

\appendix

\section{Contraction Tensors}

Contraction of tensors becomes a tedious operation as their rank becomes
larger and larger. Already at the rank-3 level, there arise various possibilities in
contracting the indices. Indeed, if one defines
\begin{eqnarray}
\label{contractx}
\mathbb{A}\bullet \mathbb{B} \equiv \mathbb{A}^{\lambda}_{\alpha\beta} \Xi^{\alpha\beta\mu\nu}_{\lambda\rho}
\mathbb{B}^{\rho}_{\mu\nu}
\end{eqnarray}
the contraction tensor $\Xi^{\alpha\beta\mu\nu}_{\lambda\rho}$ is found to have the most general form
\begin{eqnarray}
\Xi^{\alpha\beta\mu\nu}_{\lambda\rho} &=& g_{\lambda\rho}\left(g^{\alpha\beta} g^{\mu\nu} \oplus g^{\alpha\mu} g^{\beta
\nu} \oplus g^{\alpha\nu} g^{\beta \mu}\right)\nonumber\\
&\oplus& \delta_{\lambda}^{\alpha} \left( g^{\beta\nu} \delta_{\rho}^{\mu} \oplus g^{\beta \mu} \delta_{\rho}^{\nu}
\oplus g^{\mu \nu} \delta_{\rho}^{\beta}\right)\nonumber\\
&\oplus& \delta_{\lambda}^{\beta} \left( g^{\alpha\nu} \delta_{\rho}^{\mu} \oplus g^{\alpha \mu} \delta_{\rho}^{\nu}
\oplus g^{\mu \nu} \delta_{\rho}^{\alpha}\right)\nonumber\\
&\oplus& \delta_{\lambda}^{\mu} \left( g^{\beta \nu} \delta_{\rho}^{\alpha} \oplus g^{\alpha\beta} \delta_{\rho}^{\nu}
\oplus  g^{\alpha \nu} \delta_{\rho}^{\beta}  \right)\nonumber\\
&\oplus& \delta_{\lambda}^{\nu} \left( g^{\beta \mu} \delta_{\rho}^{\alpha} \oplus g^{\alpha\beta} \delta_{\rho}^{\mu}
\oplus  g^{\alpha \mu} \delta_{\rho}^{\beta}  \right)
\end{eqnarray}
where $\oplus$ implies $+$ or $-$ depending on whether symmetric or antisymmetric combinations of the indices are
involved. Clearly, $\oplus$ also contains
the appropriate symmetry factors.

As an example, let us take $\mathbb{B}^{\rho}_{\mu \nu} = \mathbb{S}^{\rho}_{\mu\nu}$ which is antisymmetric in
$(\mu,\nu)$. In this case, when contracting
$\mathbb{S}^{\rho}_{\mu\nu}$ with  $\Xi^{\alpha\beta\mu\nu}_{\lambda\rho}$  only the anti symmetric part of
$\Xi^{\alpha\beta\mu\nu}_{\lambda\rho}$ in $(\mu,\nu)$ matters. In other words, when $\mathbb{B}^{\rho}_{\mu \nu} =
\mathbb{S}^{\rho}_{\mu\nu}$  we consider only
\begin{eqnarray}
\label{antisym-munu}
\Xi^{\alpha\beta[\mu\nu]}_{\lambda\rho} &=& \frac{1}{2} \Bigg[ g_{\lambda\rho}\left(g^{\alpha\mu} g^{\beta \nu} -
g^{\alpha\nu} g^{\beta \mu}\right)\nonumber\\ &\oplus& \delta_{\lambda}^{\alpha} \left( g^{\beta\nu}
\delta_{\rho}^{\mu} - g^{\beta \mu} \delta_{\rho}^{\nu} \right)\nonumber\\ &\oplus& \delta_{\lambda}^{\beta} \left(
g^{\alpha\nu} \delta_{\rho}^{\mu} - g^{\alpha \mu} \delta_{\rho}^{\nu} \right) \nonumber\\
&\oplus& \Big[\delta_{\lambda}^{\mu} \left( g^{\beta \nu} \delta_{\rho}^{\alpha} \oplus g^{\alpha\beta}
\delta_{\rho}^{\nu} \oplus  g^{\alpha \nu} \delta_{\rho}^{\beta}  \right)\nonumber\\
&-& \delta_{\lambda}^{\nu} \left( g^{\beta \mu} \delta_{\rho}^{\alpha} \oplus g^{\alpha\beta} \delta_{\rho}^{\mu}
\oplus  g^{\alpha \mu} \delta_{\rho}^{\beta}  \right)\Big]\Bigg]
\end{eqnarray}
which is anti-symmetric in $(\mu, \nu)$.

If $\mathbb{A}^{\lambda}_{\alpha\beta}$ in (\ref{contractx}) is antisymmetric in $(\alpha,\beta)$ then we consider
antisymmetric part of (\ref{antisym-munu}).
\begin{eqnarray}
\label{antisym-munu-alfabeta}
\Xi^{[\alpha\beta][\mu\nu]}_{\lambda\rho} &=& \frac{1}{4} \Big[\delta_{\lambda}^{\alpha} \left( g^{\beta\nu}
\delta_{\rho}^{\mu} - g^{\beta \mu} \delta_{\rho}^{\nu} \right) - \delta_{\lambda}^{\beta} \left( g^{\alpha\nu}
\delta_{\rho}^{\mu} - g^{\alpha \mu} \delta_{\rho}^{\nu} \right) \Big]\nonumber\\
&+& \frac{1}{4}\Big[\delta_{\lambda}^{\mu} \left( g^{\beta \nu} \delta_{\rho}^{\alpha} -  g^{\alpha \nu}
\delta_{\rho}^{\beta}  \right)
- \delta_{\lambda}^{\nu} \left( g^{\beta \mu} \delta_{\rho}^{\alpha}    -   g^{\alpha \mu} \delta_{\rho}^{\beta}
\right)\Big]\nonumber\\
&+& \frac{1}{2} g_{\lambda\rho}\left(g^{\alpha\mu} g^{\beta \nu} - g^{\alpha\nu} g^{\beta \mu}\right)
\end{eqnarray}

For instance, $\mathbb{S} \bullet \mathbb{S}$ will be computed by using this contraction tensor.

However, if $\mathbb{A}^{\lambda}_{\alpha\beta}$ in (\ref{contractx}) is symmetric in $(\alpha,\beta)$ then we have to
consider symmetric part of (\ref{antisym-munu}).
\begin{eqnarray}
\label{antisym-munu-sym-alfabeta}
\Xi^{(\alpha\beta)[\mu\nu]}_{\lambda\rho} &=& \frac{1}{4} \Big[\delta_{\lambda}^{\alpha} \left( g^{\beta\nu}
\delta_{\rho}^{\mu} - g^{\beta \mu} \delta_{\rho}^{\nu} \right) + \delta_{\lambda}^{\beta} \left( g^{\alpha\nu}
\delta_{\rho}^{\mu} - g^{\alpha \mu} \delta_{\rho}^{\nu} \right) \Big]\nonumber\\
&+& \frac{1}{4}\Big[\delta_{\lambda}^{\mu} \left( g^{\beta \nu} \delta_{\rho}^{\alpha} + g^{\alpha \nu}
\delta_{\rho}^{\beta}  \right)
- \delta_{\lambda}^{\nu} \left( g^{\beta \mu} \delta_{\rho}^{\alpha}   +   g^{\alpha \mu} \delta_{\rho}^{\beta}
\right)\Big]\nonumber\\
&+& \frac{1}{2} g^{\alpha\beta} \left( \delta_{\lambda}^{\mu} \delta_{\rho}^{\nu} -  \delta_{\lambda}^{\nu}
\delta_{\rho}^{\mu}\right)
\end{eqnarray}
For instance, $\mathbb{Q}\bullet \mathbb{S}$ should be computed by
using this contraction tensor. In computing
$\mathbb{Q}\bullet\mathbb{Q}$ we should symmetrize in both
$(\mu,\nu)$ and $(\alpha,\beta)$. Then contraction tensor of
$\mathbb{Q}\bullet\mathbb{Q}$ is given
\begin{eqnarray}
\label{sym-munu-sym-alfabeta}
\Xi^{(\alpha\beta)(\mu\nu)}_{\lambda\rho}&=&g^{\alpha\beta}g^{\mu\nu}g_{\lambda\rho}+\frac{1}{2}\Big[g_{\lambda\rho}\left(g^{\alpha\mu}
g^{\beta\nu}+g^{\alpha\nu}g^{\beta\mu}\right)\nonumber\\
&+&g^{\mu\nu}\left(\delta^{\alpha}_{\lambda}
\delta^{\beta}_{\rho}+\delta^{\beta}_{\lambda} \delta^{\alpha}
_{\rho}\right)+ g^{\alpha\beta}\left(\delta^{\mu}_{\lambda}
\delta^{\nu}_{\rho}+\delta^{\nu}_{\lambda}\delta^{\mu}_{\rho}\right)\Big]\nonumber\\
&+& \frac{1}{4}\Big[\delta^{\alpha}_{\lambda}\left(g^{\beta\nu}
\delta^{\mu}_{\rho}+g^{\beta\mu}\delta^{\nu}_{\rho}\right)
+\delta^{\beta}_{\lambda}\left(g^{\alpha\nu}\delta^{\mu}_{\rho}
+g^{\alpha\mu}\delta^{\nu}_{\rho}\right)\Big]\nonumber\\
&+&\frac{1}{4}\Big[\delta^{\mu}_{\lambda}
\left(g^{\beta\nu}\delta^{\alpha}_{\rho}+
g^{\alpha\nu}\delta^{\beta}_{\rho}\right)+\delta^{\nu}_{\lambda}
\left(g^{\beta\mu}\delta^{\alpha}_{\rho}+g^{\alpha\mu}\delta^{\beta}_{\rho}\right)\Big]\nonumber\\
~~~~~~~~~
\end{eqnarray}
In addition to these, one can compute contraction of divergence of
tensors as
\begin{eqnarray}
\label{divergence general}
\nabla^{\between}A\bullet\nabla^{\between}B=
\nabla^{\between}_{\lambda}A^{\lambda}_{\alpha\beta}\Xi^{\alpha\beta\mu\nu}\nabla^{\between}_{\rho}B^{\rho}_{\mu\nu}
\end{eqnarray}
$\Xi^{\alpha\beta\mu\nu}$ is contraction tensor and defined in
general form as
\begin{eqnarray}
\label{general divergence}
\Xi^{\alpha\beta\mu\nu}=
g^{\alpha\beta}g^{\mu\nu}\oplus g^{\alpha\mu}g^{\beta\nu}\oplus
g^{\alpha\nu}g^{\beta\mu}
\end{eqnarray}
If A is symmetric in $(\alpha,\beta)$ and B is symmetric in
$(\mu,\nu)$ contraction tensor takes the form
\begin{eqnarray}
\label{symmetric-div}
\Xi^{(\alpha\beta)(\mu\nu)}=
g^{\alpha\beta}g^{\mu\nu}+\frac{1}{2}\left(g^{\alpha\mu}g^{\beta\nu}+g^{\alpha\nu}g^{\beta\mu}\right)
\end{eqnarray}
this contraction tensor can be used to compute
$\nabla\mathbb{Q}\bullet\nabla\mathbb{Q}$ because $\mathbb{Q}$ is
symmetric in $(\alpha\beta)$. To compute
$\nabla\mathbb{S}\bullet\nabla\mathbb{S}$, one needs contraction
tensor which is antisymmetric both couple of indices.So,
\begin{eqnarray}
\label{antisymm-div}
\Xi^{[\alpha\beta][\mu\nu]}= \frac{1}{2}\left(
g^{\alpha\mu}g^{\beta\nu}-g^{\alpha\nu}g^{\beta\mu}\right)
\end{eqnarray}
If one writes contraction tensor of
$\nabla\mathbb{Q}\bullet\nabla\mathbb{S}$, it is as;
\begin{eqnarray}
\label{symm-antisymm-div}
\Xi^{(\alpha\beta)[\mu\nu]}=0
\end{eqnarray}

{\section{Positive-Definite Mass Matrix}}

In the text, we mentioned that for a stable theory, each of the three eigenvalues must individually be positive. This leads to non-trivial constraints on the coefficients in (\ref{relatex}). In this appendix we shall discuss certain related details. The eigenvalues of (\ref{mass-mat}) follow from the cubic algebraic equation
%
\begin{eqnarray}
\label{char-eq}
-\lambda^{3}+b\lambda^{2}+c\lambda+d=0
\end{eqnarray}
where
\begin{eqnarray}
b&=&a_{VV}+a_{UU}+a_{WW}\nonumber \\
c&=&-a_{VV}a_{UU}-a_{VV}a_{WW}-a_{UU}a_{WW}\nonumber\\
&+&a^{2}_{UW}+a^{2}_{VU}+a^{2}_{VW}\nonumber \\
d&=&a_{VV}a_{UU}a_{WW}+2a_{VU}a_{UW}a_{VW}-a^{2}_{UW}a_{VV}\nonumber\\
&-&a_{VW}^{2}a_{UU}-a^{2}_{VU}a_{WW}\,.
\end{eqnarray}
The roots of  (\ref{char-eq}) must each be non-negative for guaranteeing absence of instabilities. The
analytic expressions for roots are well-known. However, the constraint equations they lead to are
too complicated to achieve specific statements about the elements of the mass matrix (\ref{mass-mat}).
Nevertheless, in a given specific problem, one can determine the allowed ranges for $a_{VV}, \cdots, a_{UW}$
at least numerically,

As an algebraically simpler case to exemplify, one can focus on the special case
of vanishing discriminant, that is, one considers
\begin{eqnarray}
\Delta=18 abcd - 4 b^{3}d + b^{2}c^{2}- 4 ac^{3}-27 a^{2} d^{2}
\end{eqnarray}
so that only two independent eigenvalues are left. Indeed, one has
\begin{eqnarray}
\lambda_{1}=-\frac{b}{3a}-\frac{2}{3a}\sqrt[3]{\frac{1}{2}[2b^{3}-9abc+27a^{2}d]}
\end{eqnarray}
and
\begin{eqnarray}
\lambda_{2}=-\frac{b}{3a}+\frac{1}{3a}\sqrt[3]{\frac{1}{2}[2b^{3}-9abc+27a^{2}d]}\,.
\end{eqnarray}
For positive-definite mass matrix, $\lambda_{1}$ and $\lambda_{2}$ must each be positive:
\begin{eqnarray}
\lambda_{1}> 0   \Rightarrow -\frac{b}{2} < \sqrt[3]{\frac{1}{2}[2b^{3}-9abc+27a^{2}d]}
\end{eqnarray}
and
\begin{eqnarray}
\lambda_{2}> 0   \Rightarrow b > \sqrt[3]{\frac{1}{2}[2b^{3}-9abc+27a^{2}d]}\,.
\end{eqnarray}
These two constraints lead one at once to the bound
\begin{eqnarray}
-\frac{b}{2} < \sqrt[3]{\frac{1}{2}[2b^{3}-9abc+27a^{2}d]} < b\,.
\end{eqnarray}
Similar bounds can be derived for general as well as special cases \cite{DAD}. In general, constraints on various coefficients become more suggestive in some physically relevant special cases. We here thus exemplify two such cases: Symmetric and Antisymmetric connections.

\begin{enumerate}

\item {Symmetric Connection} : $\between^{\lambda}_{\alpha \beta}=\between^{\lambda}_{ \beta \alpha}$ \\

As we have mentioned in the text, in this case, torsion tensor identically vanishes ($\mathbb{S}^{\lambda}_{\alpha \beta}=0$), and consequently $\texttt{V}_{\alpha}=\texttt{U}_{\alpha}$. The theory then reduces to a two-vector theory of $\texttt{V}$ and $\texttt{W}$.
From Eq. (\ref{action-acik}) the mass-squared matrix of vectors is found to be
\begin{eqnarray}
\label{mass-matt}
\frac{1}{2} M_{Pl}^2 \left(\begin{array}{cc} a'_{VV}+a'_{UU}+2 a'_{VU} \quad a'_{VW}+a'_{UW}\\
a'_{VW}+a'_{UW} \quad  a'_{WW} \end{array} \right)
\end{eqnarray}
where various coefficients are given by
\begin{eqnarray}
\label{relatexx}
a'_{VV} &=& \frac{1}{18} + \frac{44}{9} c_{Q} \,,\nonumber\\
a'_{UU} &=& a_{WW}  \,,\nonumber\\
a'_{WW} &=& \frac{1}{18} + \frac{14}{9} c_{Q}\,,\nonumber\\
a'_{VU} &=& -\frac{1}{9}  + \frac{20}{9} c_{Q} \,,\nonumber\\
a'_{VW} &=& -\frac{1}{9} + \frac{20}{9} c_{Q} \,,\nonumber\\
a'_{UW} &=& \frac{7}{18} + \frac{14}{9} c_{Q} \,.
\end{eqnarray}
which follow from (\ref{action-acik}) for vanishing torsion. Clearly,
$c_Q$ is the only variable. The eigenvalues of (\ref{mass-matt}) follow
from the quadratic algebraic equation;
\begin{eqnarray}
\label{quadric}
\lambda^2+b'\lambda+c'=0
\end{eqnarray}
where
\begin{eqnarray}
b'&=&-a_{VV}'-a_{UU}'-2a_{UU}'-a_{WW}'\nonumber \\
c'&=& (a_{VV}'+a_{UU}'+2a_{UU}')a_{WW}'- (a_{VW}'+a_{UW}')^2\,.
\end{eqnarray}
From the equation (\ref{quadric}), one directly determines the discriminant
\begin{eqnarray}
\Delta = \frac{11680}{81}c_{Q}^2+\frac{872}{162}c_{Q}+\frac{109}{324}
\end{eqnarray}
and eigenvalues
\begin{eqnarray}
\lambda_{1,2}&=&\frac{a_{VV}'+a_{UU}'+2a_{UU}'+a_{WW}'\pm\sqrt{\Delta}}{2}\nonumber\\
&=& \frac{-\frac{1}{18}+\frac{112}{9}c_{Q}\pm \sqrt{\Delta}}{2}
\end{eqnarray}
For a physically sensible theory, the eigenvalues must all be positive. By considering the constraint of positive discriminant and roots, one finds two appropriate intervals
\begin{eqnarray}
c_{Q} < - 0.046 \quad  c_{Q}> 0.68\, .
\end{eqnarray}
This shows that except for the  small interval containing origin, all values of $c_{Q}$ lead to a stable massive two-vector
theory.

\item {Anti-symmetric tensorial connection}: $V_{\alpha}=-U_{\alpha}$ and $W_{\alpha}=0$ \\

In this case we end up with a single-vector theory with mass-squared  $\frac{1}{2}M_{Pl}^2\bar{a}_{VV}$ where $\bar{a}_{VV}=1/3+8c_{S}+2c_{Q}+8c_{QS}$. This coefficient must be positive and hence
 \begin{eqnarray}
4c_{S}+c_{Q}+4c_{QS}> -\frac{1}{6}
\end{eqnarray}
A much more special arises when non-metricity vanishes. In this special case, the coefficients $c_{Q}$ and $c_{QS}$ both vanishe, and one finds
\begin{eqnarray}
c_{S}>-\frac{1}{24}
\end{eqnarray}
as a bound on $c_S$.

\end{enumerate}

\end{document}